\RequirePackage{ifpdf}
\ifpdf 
\documentclass[pdftex]{sigma}
\else
\documentclass{sigma}
\fi

\newcommand*{\be}{\begin{equation}}
\newcommand*{\ee}{\end{equation}}
\newcommand*{\ud}{\mathrm{d}}

\begin{document}

\allowdisplaybreaks

\renewcommand{\textfraction}{0.01}
\renewcommand{\topfraction}{0.99}

\renewcommand{\PaperNumber}{083}

\FirstPageHeading

\renewcommand{\thefootnote}{$\star$}

\ShortArticleName{Stability Analysis of Continuous Waves in Nonlocal Random Nonlinear Media}

\ArticleName{Stability Analysis of Continuous Waves\\ in Nonlocal Random Nonlinear Media\footnote{This paper is a
contribution to the Proceedings of the Seventh International Conference ``Symmetry in Nonlinear Mathematical Physics''
(June 24--30, 2007, Kyiv, Ukraine). The full collection is available at
\href{http://www.emis.de/journals/SIGMA/symmetry2007.html}{http://www.emis.de/journals/SIGMA/symmetry2007.html}}}

\Author{Maxim A. MOLCHAN}

\AuthorNameForHeading{M.A. Molchan}

\Address{B.I. Stepanov Institute of Physics, 68
Nezalezhnasci Ave., 220072 Minsk, Belarus}
\Email{\href{mailto:m.moltschan@dragon.bas-net.by}{m.moltschan@dragon.bas-net.by}}

\ArticleDates{Received July 26, 2007, in f\/inal form August 15, 2007; Published online August 26, 2007}

\Abstract{On the basis of the competing cubic-quintic nonlinearity model, stability (instability) of continuous waves in nonlocal random non-Kerr nonlinear media is studied analy\-ti\-cally and numerically. Fluctuating media parameters are modeled by the Gaussian white noise. It is shown that for dif\/ferent response functions of a medium nonlocality suppresses, as a rule, both the growth rate peak and bandwidth of instability caused by random parameters. At the same time, for a special form of the response functions there can be an ``anomalous'' subjection of nonlocality to the instability development which leads to further increase of the growth rate. Along with the second-order moments of the modulational amplitude, higher-order moments are taken into account.}

\Keywords{nonlocality; competing nonlinearity; stochasticity}

\Classification{78A10; 45K05}

\section{Introduction}

\label{sect:intro}
Modulational instability (MI) is a universal
phenomenon arising in many nonlinear systems. It is closely
related to formation of localized wave structures -- solitons and
considered as a~precursor of this formation. In optics, in case of
Kerr-like nonlinear media MI is responsible for formation of
bright solitons and its absence is necessary for dark solitons
formation~\cite{Rev1}. MI consisting in the exponential growth of
small perturbations in the amplitude or the phase of initial
optical waves leads to the breakup of a beam or a quasi-coherent
pulse into f\/ilaments that can evolve into trains of
solitons~\cite{Rev2}. MI f\/inds its application in
plasma~\cite{Rev3}, hydrodynamics~\cite{Rev4}, f\/luids~\cite{Rev5},
atomic Bose--Einstein condensates~\cite{Rev6,Rev7}. The realm of
applications also includes optical communications systems,
all-optical logical devices~\cite{Rev8}, and the generation of
ultra-high repetition-rate trains of soliton-like
pulses~\cite{Rev9}.

In the setting of nonlinear optics it was shown that MI is absent for defocusing Kerr-type media, the medium being described by deterministic parameters. In case of focusing deterministic Kerr media MI presents as the long-wave instability with a f\/inite bandwidth~\cite{Rev10}. In frames of a more realistic approach that implies random behavior of the characteristic parameters of a medium, that is, they are considered to f\/luctuate randomly around their mean values, the region of MI was shown to be extended by stochastic inhomogeneities of a Kerr medium over the whole spectrum of modulation wave numbers. This is true for both focusing and defocusing regimes~\cite{Rev11}.

The next step toward more realistic description implies taking into account nonlocality of a medium. Moreover, some media manifestly exhibit nonlocal properties~\cite{Rev12} and should be described by appropriate nonlocal models. As a rule, this property is a result of a corresponding transport process such as heat conduction
in thermal nonlinear media~\cite{Rev13}, dif\/fusion of atoms in a gas~\cite{Rev14}, many-body interaction in
Bose--Einstein condensates~\cite{Rev15}. Nonlocality, as compared to local cases, can crucially change the behavior of proceeding in such a medium processes. For example, it can support the propagation of self-focused beams without collapsing~\cite{Rev16, Rev16a}. Dark solitons interaction~\cite{Rev17} is greatly inf\/luenced by nonlocal properties of a medium. The investigation of MI in deterministic nonlocal Kerr-type media revealed its absence for the defocusing case for small and moderate values of the product ``modulation amplitude $\times$ nonlocality parameter"~\cite{Rev18}. The overview of the modulational instability and beam propagation in nonlocal nonlinear media can be found in~\cite{Rev18a}.

The incorporation of nonlocality into a model under study is generally accomplished via the nonlinear refractive index of a material, whose dependence on the intensity is determined by the chosen model of nonlinearity.
Usually, experiments show the deviation from the linear (Kerr) dependence of the refractive index on the incident intensity for large intensities. This occurs for a range of materials: semiconductor-doped glasses~\cite{Rev19}, semiconductor waveguides~\cite{Rev20}, and nonlinear polymers~\cite{Rev21}. Saturable and cubic-quintic nonlinearities are the most typical for nonlinear systems.

The objective of this paper is to investigate MI in nonlocal medium with stochastic parameters
and competing cubic-quintic nonlinearity.
The fact that nonlocality spreads out localized excitations allows to anticipate the decrease of both the growth rate peaks and bandwidths of instability. Such a situation is valid for the case of stochastic media with the sign-def\/inite Fourier images of the response functions. However, for nonlocal media with sign-indef\/inite Fourier image of the response function MI gain can exceed that of a local stochastic medium.
This behavior can be considered as ``anomalous''.

In the present paper the investigation of MI of continuous waves in nonlocal stochastic media is based on a generalized nonlocal nonlinear Schr\"odinger equation with random coef\/f\/icients and competing nonlinearities.
For the illustration of the obtained results we use the white noise model for parameter f\/luctuations and
response functions of several types.

\section{Model}

 In this paper we consider a medium with nonlinearity that is a nonlocal function of the incident f\/ield. This nonlinearity induced by a wave with the intensity $I(x,z)$ can be presented in general form
\be\label{Rindex}
\Delta n(I)=g(z)\int_{-\infty}^\infty\ud x'R(x-x')I^a(x',z)+s(z)\int_{-\infty}^\infty\ud x'L(x-x')I^{2a}(x',z),
\ee
where $x$ is the transverse coordinate, $a$ is a positive constant, $z$ can be considered as time or the spatial propagation coordinate depending on the context, nonlinearity coef\/f\/icients
$g(z)$ and $s(z)$ are considered as stochastic functions which f\/luctuate
around their mean values $g_{0}$ and $s_{0}$:
\be\label{d(z)}
g(z)=g_0(1+m_g(z)),\qquad s(z)=s_0(1+m_s(z)),
\ee
Here $m_{g}$ and $m_{s}$ are independent zero-mean random
processes of the Gaussian white-noise type,
\begin{gather*}
\langle m_g\rangle=\langle m_s\rangle=0,\qquad
\langle m_g(z)m_g(z')\rangle=2\sigma_g^{2}\delta(z-z'),
\qquad
\langle m_s(z)m_s(z')\rangle=2\sigma_s^{2}\delta(z-z').
\end{gather*}
The angle brackets designate the expectation with respect to
the distribution of the correspon\-ding processes.
In the imposed phenomenological model~(\ref{Rindex}) the
f\/ield-intensity dependent change of the refractive index is characterized by two
normalized symmetric response functions $R(x)$ and $L(x)$,
$\int_{-\infty}^\infty\ud xR(x)=\int_{-\infty}^\infty\ud xL(x)=1$.
In case when $R(x)=L(x)=\delta(x)$ (local limit) and constant nonlinearity coef\/f\/icients we reproduce the well-known general power-law dependence on the incident intensity for local models with competing nonlinearities~\cite{Rev1}. At the same time, for equal response functions one can consider the expression as the second-order expansion of the model with saturable nonlinearity, provided $a=1$ and $s=-g/I_s=-n_\infty/I_{s}^{2}$, where $I_s$ is the saturation intensity and $n_\infty$ corresponds to the maximum change in the refractive index. The set of the parameters $a=1$ and $s(z)=0$ is associated with nonlocal Kerr-type media with random parameters~\cite{Rev22}.

Thus the propagation of plane waves along the $z$ axis in a stochastic medium with nonlocal competing nonlinearities is governed by the generalized nonlinear Schr\"odinger equation
\be\label{Ceq}
iu_z+\frac{1}{2}d(z)u_{xx}+u\Delta n(|u|^2)=0,
\ee
where $u(x,z)$ is the complex envelope amplitude, $\Delta
n(|u|^2)$ is def\/ined by~(\ref{Rindex}) and standard dimensionless
variables are used. The dif\/fraction coef\/f\/icient $d(z)$, just as
$g(z)$ and $s(z)$, is a stochastic function with the same
properties:
\[
\langle m_d\rangle=0,\qquad
\langle m_d(z)m_d(z')\rangle=2\sigma_d^{2}\delta(z-z').
\]
Equation (\ref{Ceq}) represents the generalization of the Kerr-type nonlocal model considered in \cite{Rev22} for the case of higher-order nonlinearities. The deterministic variant was comprehensively studied in \cite{Rev22a}.

In further considerations we adopt that $d_0>0$ and the condition $g_0\cdot s_0<0$ stands for the competition of nonlinearities.

As a solution equation ({\ref{Ceq}}) admits a homogeneous plane wave
\be\label{eq4}
u_{0}=A\exp\left[iA^{2a} \int_0^z\ud z' \left(g(z')+A^{2a}s(z')\right)\right],
\ee
where $A$ is a real amplitude. Proceeding to a linear
stability analysis of the solution~(\ref{eq4}) we assume that
\be\label{eq5}
u(x,z)=\left(A+v(x,z)\right)\exp\left[iA^{2a} \int_0^z\ud z' \left(g(z')+A^{2a}s(z')\right)\right]
\ee
is a perturbed solution of equation~(\ref{Ceq}). Here $v(x,z)$ is a small complex modulation.
The substitution of equation~(\ref{eq5}) into equation~(\ref{Ceq}) and the linearization
around the plane wave~(\ref{eq4}) provide a linear equation for
$v(x,z)$:
\begin{gather}
iv_{z}+\frac{1}{2}d(z)v_{xx}+2ag(z)A^{2a}\int \ud x'R(x-x')
\textrm{Re}\,v(x',z)\nonumber\\
\qquad{}+4as(z)A^{4a}\int \ud x'L(x-x')
\textrm{Re}\,v(x',z)=0.\label{eq6}
\end{gather}

Decomposing $v$ into real and
imaginary parts, $v=v_r(x,z)+iv_i(x,z)$, and performing the Fourier
transforms
\begin{gather*}
\rho(k,z)=\frac{1}{2\pi}\int_{-\infty}^\infty\ud x\,v_r(x,z)e^{ikx},
\qquad
\sigma(k,z)=\frac{1}{2\pi}\int_{-\infty}^\infty\ud x\,v_i(x,z)e^{ikx},
\\
\hat{R}(k,z)=\frac{1}{2\pi}\int_{-\infty}^\infty\ud x\,R(x,z)e^{ikx},
\qquad
\hat{L}(k,z)=\frac{1}{2\pi}\int_{-\infty}^\infty\ud x\,L(x,z)e^{ikx},
\end{gather*}
we convert
equation~(\ref{eq6}) to a system of
linear equations for $\rho$ and $\sigma$:
\be\label{eq7}
\frac{\ud}{\ud z}\!\! \left( \begin{array}{ccc}
\rho \\
\sigma \\
\end{array} \right)
=
\left( \begin{array}{ccc}
0 & \frac{1}{2}d(z)k^2 \\
-\frac{1}{2}d(z)k^2+2ag(z)A^{2a}\hat R +4as(z)A^{4a}\hat{L}& 0 \\
\end{array} \right)
\left( \begin{array}{ccc}
\rho \\
\sigma \\
\end{array} \right).
\ee

Equation~(\ref{eq7}) is suf\/f\/icient for the investigation of MI for a
deterministic medium with the parameters $d_0$, $g_0$, and $s_0$.
In contrast, for stochastic systems one should consider the second
moments $\langle\rho^2\rangle$, $\langle\rho\sigma\rangle$ and
$\langle\sigma^2\rangle$ providing the minimal nontrivial
information about MI induced by random f\/luctuations.

\section{The second-order moment MI gain}
\subsection{Theoretical approach}
We consider the vector of the second moments
\be \label{eq8}
X^{(2)}=\left(\langle\rho^2\rangle,\langle\rho\sigma\rangle,\langle\sigma^2\rangle\right)^T.
\ee
The Furutsu--Novikov formula~\cite{Rev23}
\be\label{FNf}
\langle u_i(\textbf{r})F[\textbf{r}]\rangle=\sum_{j}\int\ud \textbf{r}'\,B_{ij}(\textbf{r}-\textbf{r}')\left< \frac{\delta F[\textbf{u}]}{\delta u_{j}(\textbf{r}')}
\right>,
\ee
where $\textbf{u}$ is an arbitrary zero-mean Gaussian f\/ield with the covariance function $B_{ij}(\textbf{r})$, $F[\textbf{u}]$ is a~functional, allows to express products like $\langle m_d\rho^2\rangle$ in terms of components of $X^{(2)}$. As a result, $X^{(2)}$ is subject to the evolution equation $\left(\ud/\ud z\right)X^{(2)}=M^{(2)}X^{(2)}$, where $M^{(2)}$ is the $3\times 3$
matrix of the form
\be
\label{eq10}
M^{(2)}=\left(\begin{array}{ccc}
{-\frac{1}{2}A} & {B} &
\frac{1}{2}A\vspace{1mm} \\
{C-\frac{1}{2}B} &
{-A} & {\frac{1}{2}B} \vspace{1mm}\\
{\frac{1}{2}\left(D+A\right)} & {2\left(C-\frac{1}{2}B\right)}\; &
{-\frac{1}{2}A}
\end{array}\right),
\ee
where
\begin{gather*}
A=\sigma_d^2d_0^2k^4,
\qquad
B={d_{0}k^2},
\qquad
C={2ag_0A^{2a}\hat R+4as_0A^{4a}\hat L},
\\
D={64a^2\sigma_{s}^{2}s_0^2A^{8a}\hat L^2+ 16a^2\sigma_{g}^{2}g_{0}^{2}A^{4a}\hat R^2}.
\end{gather*}
Instability occurs for positive real parts of the eigenvalues of $M^{(2)}$.
MI gain $G_{2}(k)$ is determined by the largest
positive value. We will illustrate the results by means of the exponential response function with the sign-def\/inite Fourier image
\be\label{exp}
R_e(x)=\frac{1}{2\lambda}\exp\left( -\frac{|x|}{\lambda}\right),\qquad
\hat{R}_e(k)=\frac{1}{1+\lambda^2k^2}
\ee
and as an example of the sign-indef\/inite Fourier transform response function we will take the
rectangular response function
\be\label{sin}
R_r(x)=\left\{\begin{array}{cc} \displaystyle{\frac{1}{2\lambda}} \;&
\mathrm{for} \; |x|\le \lambda,\vspace{2mm}\\ 0 \;& \mathrm{for} \;
|x|>\lambda,
\end{array}\right. \qquad \hat R_r(k)=\frac{\sin(\lambda k)}{\lambda k},
\ee
where $\lambda$ is the nonlocality
parameter. Below we will study in detail the case of competing cubic-quintic nonlinearity, that is,
$a=1$, $g_0>0$ and $s_0<0$ and the response functions are assumed to be equal. For analysis purposes it is convenient to introduce a quantity
\[
\chi=\frac{g_0}{2|s_0|A^2}
\]
with the help of what we will discriminate between the focusing and the defocusing regimes. The inequality $\chi>1$ corresponds to the predominance of the cubic nonlinearity term over the defocusing one, $\chi<1$ determines the defocusing regime when the quintic term prevails.

\begin{figure}[t]\small
\centering
\begin{tabular}{@{}ccc@{}}
\includegraphics[height=38mm]{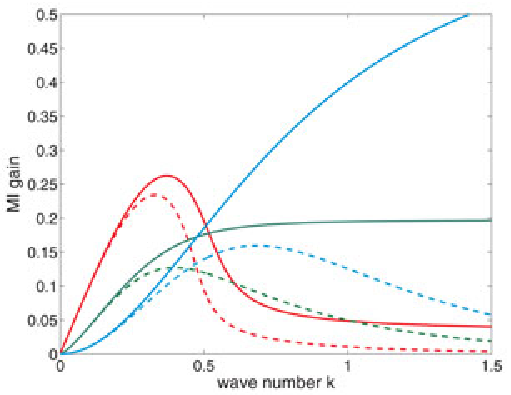} &
\includegraphics[height=38mm]{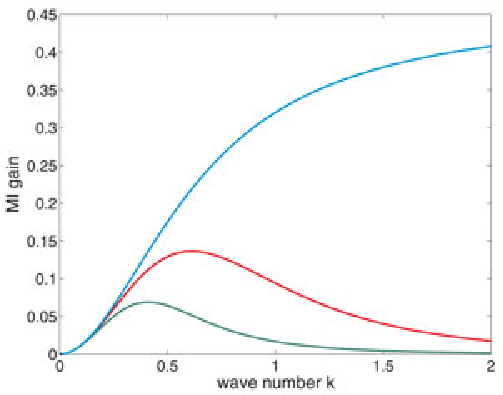} &
\includegraphics[height=38mm]{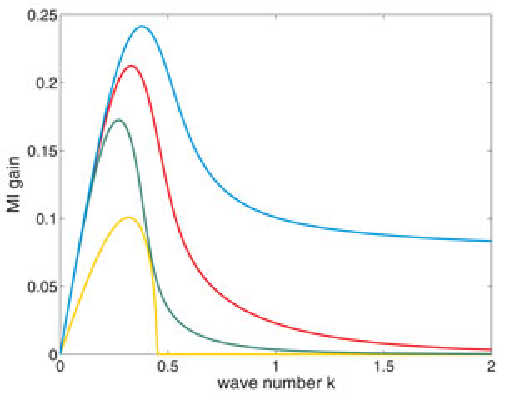}\\
(a) & (b) & (c)
\end{tabular}
\caption{Plots of the MI gain $G_{2}(k)$ for: (a) local stochastic media (solid lines) and nonlocal stochastic media with the exponential response function (dashed lines) for dif\/ferent values of the parameter $\chi$: the pair of red curves corresponds to $\chi=2$, blue ones -- to $\chi=1$, green ones -- to $\chi=0.8$, here $\lambda=1$; (b): local stochastic media (blue line) and nonlocal stochastic media with the exponential response function for $\chi=0.8$: $\lambda^2=1$ (red line), $\lambda^2=18$ (green line); (c): local deterministic media with competing cubic-quintic nonlinearity (yellow line), local stochastic media (blue line), and nonlocal stochastic media with the exponential response function for $\chi=1.4$: $\lambda^2=1$ (red line), $\lambda^2=18$ (green line). On all plots: $d_{0}=2$, $g_{0}A^2=1$, $\sigma_d^2=\sigma_g^2=\sigma_s^2=0.1$.}
\label{fig1}
\end{figure}

\begin{figure}[t]\small
\centering
\begin{tabular}{@{}ccc@{}}
\includegraphics[height=55mm]{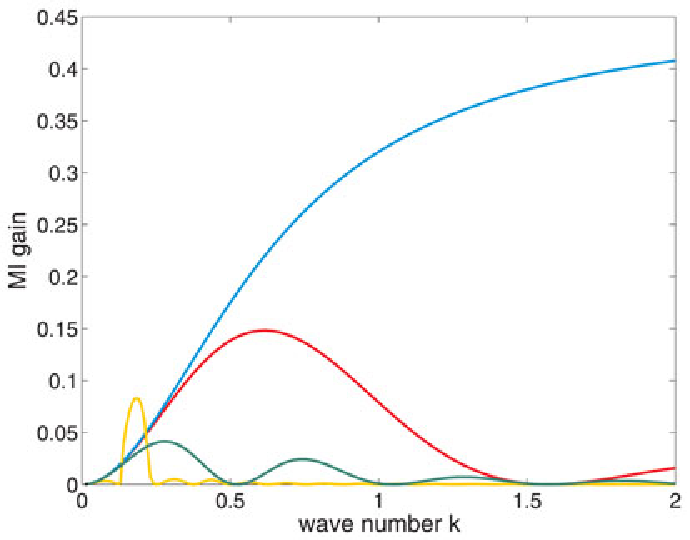}&&
\includegraphics[height=55mm]{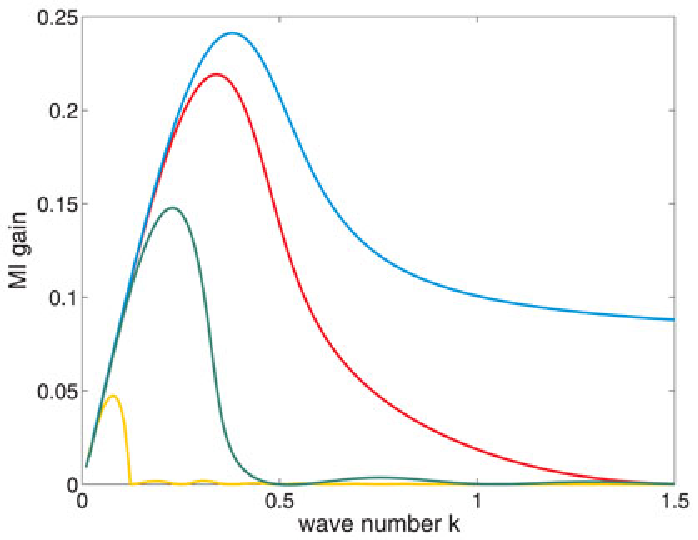}\\
(a) && (b)
\end{tabular}
\caption{Plots of the MI gain
$G_{2}(k)$ for local stochastic
media (blue lines) and nonlocal
stochastic media with the rectangular response function: (a) $\chi=0.8$, (b) $\chi=1.4$. On both plots $\lambda=2$ (red lines), $\lambda=6$ (green lines), $\lambda=25$ (yellow lines), $d_{0}=2$, $g_{0}A^2=1$, $\sigma_d^2=\sigma_g^2=\sigma_s^2=0.1$.}\label{fig2}
\end{figure}

\subsection{Numerical results}

In Fig.~\ref{fig1}(a) we demonstrate the inf\/luence of the transition from the focusing regime ($\chi>1$) to the defocusing one ($\chi<1$) on the MI bandwidth and the MI gain peak for media with cubic-quintic nonlinearity as the intensity of the incident f\/ield is increased. This transition is accompanied by the pronounced enhancement of the MI bandwidth. At the same time the MI gain peak is decreased.
For the defocusing regime the matrix (\ref{eq10}) possesses one real
eigenvalue~$\lambda_1$ that is positive for all~$k$ and two complex conjugate ones~$\lambda_2$ and
$\lambda_3$ with negative real parts in the case of the exponential response function.
The growth rate peak of $G_2(k)\equiv
\lambda_1$ and MI bandwidth are suppressed by nonlocality. The growth of the nonlocality parameter $\lambda$ results in the
amplif\/ication of the suppression ef\/fect (Fig.~\ref{fig1}(b)). The
situation for the rectangular response function (\ref{sin})
is dif\/ferent. For suf\/f\/iciently high nonlocality MI gain maximum for a given wave number $k$ can exceed the corresponding value of $G_2$ for a local random medium with cubic-quintic nonlinearity (left
panel of Fig.~\ref{fig2}). At the same time, the MI bandwidth becomes
strictly f\/inite in this limit
A local deterministic medium with the competing cubic-quintic nonlocality in the focusing regime $(\chi>1)$ (Fig. \ref{fig1}(c)) produces the long-wave instability
with a f\/inite bandwidth. The bandwidth is extended by
stochasticity of medium parameters to the whole spectrum of
modulation wave numbers. Corresponding calculation of eigenvalues
of the mat\-rix~$M_2$~(\ref{eq10}) for the case of $\chi>1$ demonstrates the
suppression the MI gain and bandwidth for media with both
sigh-def\/inite and sign-indef\/inite response functions~(Fig.~\ref{fig2}(b)). The positions of
the MI gains shift toward smaller wave numbers $k$ under
nonlocality growth, producing f\/inite bandwidth.

\section{Higher-order moments}
\subsection{Theoretical approach}
Evidently, in frames of the second-order moments analysis~(\ref{eq8}) it is impossible to characterize in full detail the dynamics of a stochastic system. For example, featured points of MI gain (maximum position, bandwidth, etc.) calculated from the second moments could f\/luctuate when accounting for higher-order moments. More detailed information about MI is provided by the higher-order moments
\be \label{eq11}
X^{(2n)}=\big
\{\langle \rho^{(2n-j)}\sigma^{j}\rangle\big\},\qquad j=0,\ldots , 2n.
\ee
As before, applying the Furutsu--Novikov
formula, we obtain a matrix $M^{(2n)}$ in the form
\begin{gather}
M^{(2n)}=d_{0}k^2 A^{(2n)}+\big(2ag_{0}A^{2a}\hat R+4as_0A^{4a}\hat L -\tfrac{1}{2}d_{0}k^2\big)B^{(2n)}
\nonumber\\
\phantom{M^{(2n)}=}{} +d_{0}^{2}k^4\sigma_{d}^{2}C^{(2n)}+\big(64a^2s_0^2A^{8a}\hat L\sigma_{s}^2+16a^2g_{0}^{2}A^{4a}\hat R^2\sigma_{g}^{2}\big) D^{(2n)}\label{eq}
\end{gather}
with the following non-zero entries of the matrices $A^{(2n)}$, $B^{(2n)}$, $C^{(2n)}$ and $D^{(2n)}$
\begin{gather*}
A_{j,j+1}^{(2n)}=n-\frac{j}{2},\qquad B_{j,j-1}^{(2n)}=j, \qquad
C_{jj}^{(2n)}=-\frac{1}{2}(n+2nj-j^2),
\nonumber\\
C_{j,j+2}^{(2n)}=\left(n-\frac{j}{2}\right)
\left(n-\frac{j+1}{2}\right),\qquad C_{j,j-2}^{(2n)}=D_{j,j-2}^{(2n)}=\frac{1}{4}j(j-1),\qquad
j=0,\ldots, 2n.
\end{gather*}

Then among the roots of the characteristic
polynomial $\det|M^{(2n)}-\lambda I|$ the maximal real part will determine $nG_{2n}(k)$. The fact that the characteristic polynomial is of the odd degree and all
the matrix elements of $M^{(2n)}$ are real ensures at least one real eigenvalue of
$M^{(2n)}$, the others being mutually complex conjugate. Below the analysis of the 4-th and 6-th moments is given.

\begin{figure}[t]\small
\centering
\begin{tabular}{@{}ccc@{}}
\includegraphics[height=55mm]{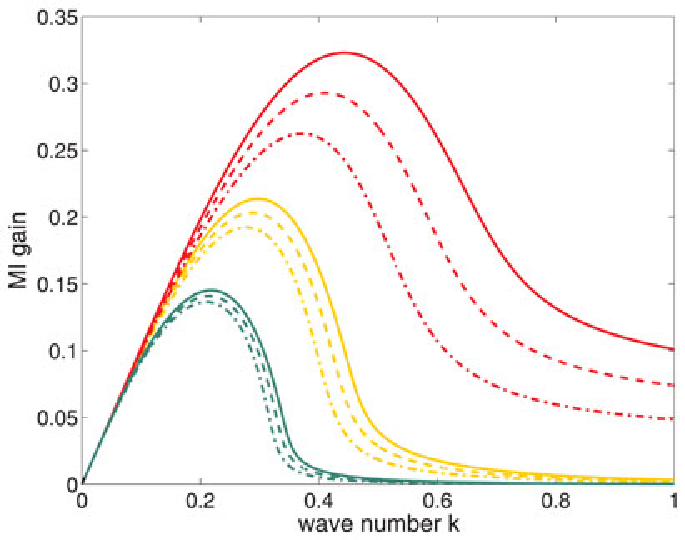} &&
\includegraphics[height=55mm]{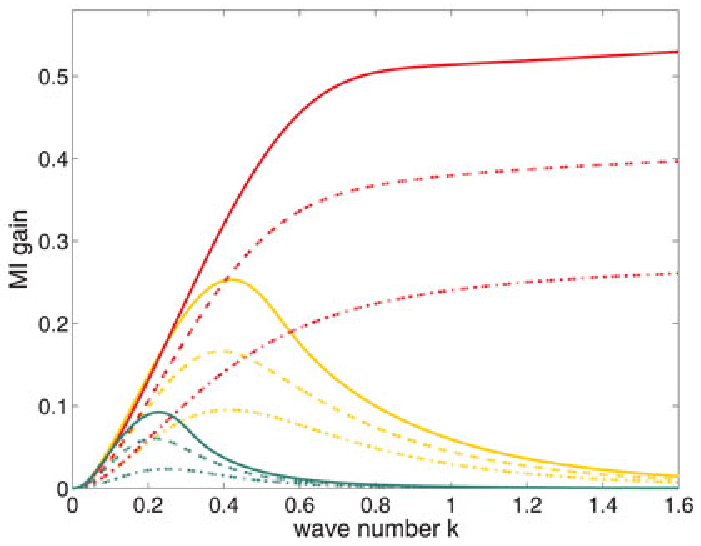}\\
(a) && (b)
\end{tabular}
\caption{Plots of the MI gains
$G_{6}(k)$ (solid lines), $G_{4}$ (dashed lines), and $G_{2}$
(dash-dotted lines) for local stochastic
media (red curves on both plots) and nonlocal
stochastic media with the exponential response function; (a): $\chi=2$, $\lambda=2$ (yellow curves), $\lambda=4$ (green curves); (b): $\chi=0.9$, $\lambda=\sqrt{2}$ (yellow curves), $\lambda=4$ (green curves). On both plots $d_{0}=2$, $g_{0}A^2=1$, $\sigma_d^2=\sigma_g^2=\sigma_s^2=0.1$.} \label{fig3}
\end{figure}

\begin{figure}[t]\small
\centering
\begin{tabular}{@{}ccc@{}}
\includegraphics[height=55mm]{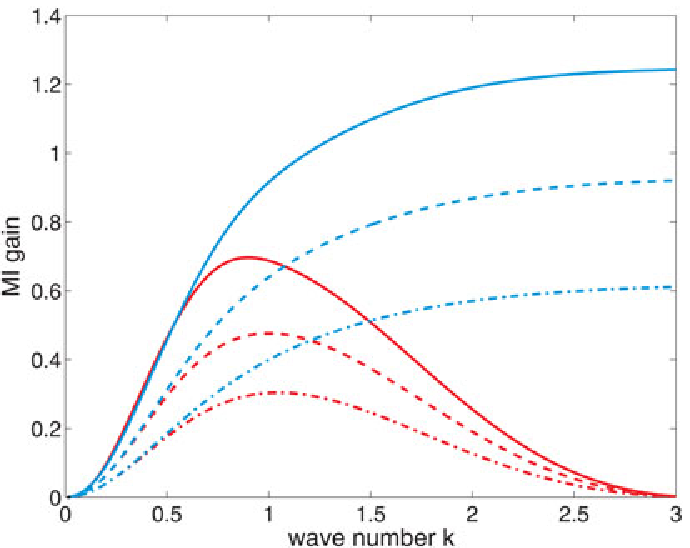}&&
\includegraphics[height=55mm]{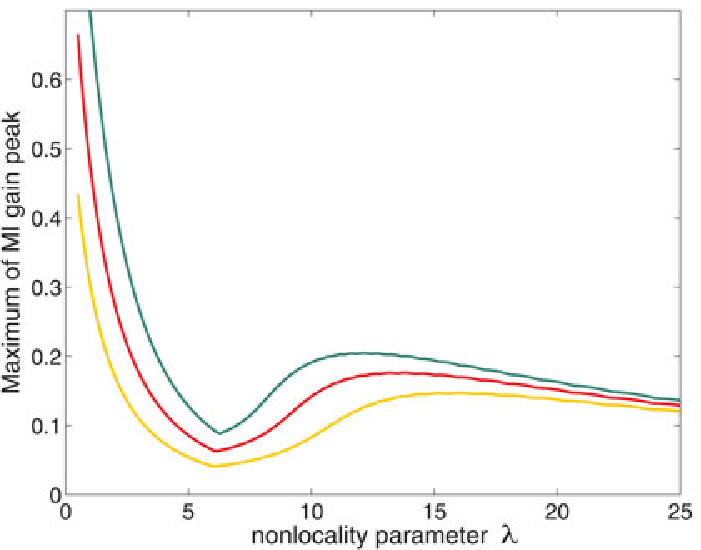}\\
(a) & & (b)
\end{tabular}
\caption{(a) Plots of the MI gains
$G_{6}(k)$ (solid lines), $G_{4}$ (dashed lines), and $G_{2}$
(dash-dotted lines) for local stochastic media (blue curves) and nonlocal stochastic media with the rectangular response function (red curves); here $\lambda=1$. (b) Plots of the maximum of the main peak of the MI gains
$G_{6}(k)$ (green line), $G_{4}$ (red line), and $G_{2}$
(yellow line) versus nonlocality parameter $\lambda$. On both plots $\chi=0.8$, $d_{0}=2$, $g_{0}A^2=1$, $\sigma_d^2=\sigma_g^2=\sigma_s^2=0.1$.} \label{fig4}
\end{figure}

\begin{figure}[t]\small
\centering
\begin{tabular}{@{}ccc@{}}
\includegraphics[height=55mm]{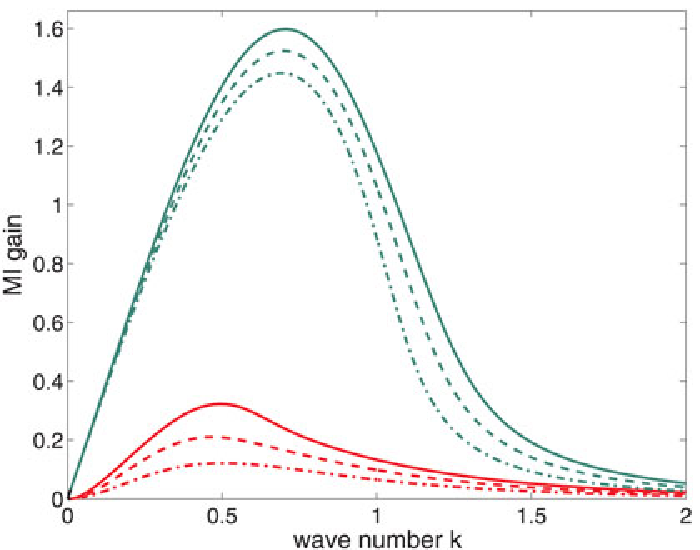} &&
\includegraphics[height=55mm]{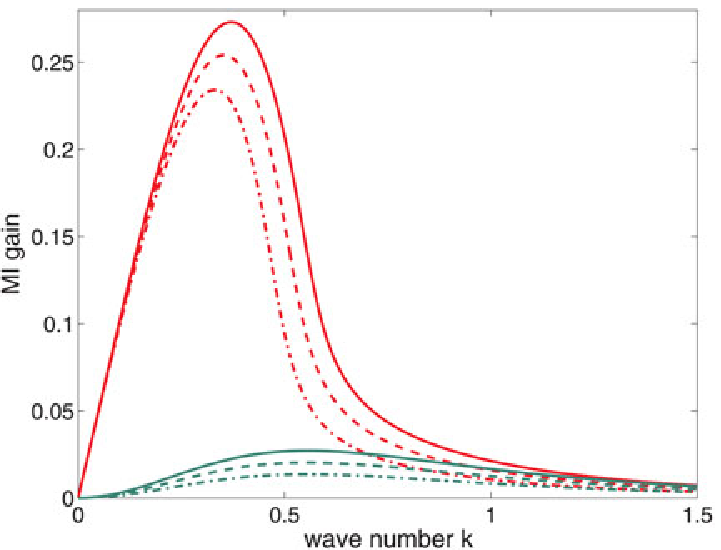}\\
(a) &&  (b)
\end{tabular}
\caption{Plots of the MI gains
$G_{6}(k)$ (solid lines), $G_{4}$ (dashed lines), and $G_{2}$
(dash-dotted lines) for nonlocal
stochastic media with the exponential response function for competing cubic-quintic nonlinearity with $g_0>0$ and $s_0<0$ (red curves on both plots) and: (a) $g_0>0$ and $s_0>0$ (green curves), here $\chi=0.9$; (b) $g_0<0$ and $s_0>0$ (green curves), here $\chi=2$. On both plots $d_{0}=2$, $g_{0}A^2=1$, $\sigma_d^2=\sigma_g^2=\sigma_s^2=0.1$, $\lambda=1$.} \label{fig5}
\end{figure}

\subsection{Numerical results}
Fig.~\ref{fig3} presents the results of calculating MI gains
$G_2$, $G_4$ and $G_6$ for the exponential response function. In case of the focusing regime as well for the defocusing one nonlocality suppresses the higher-order moments. For the focusing regime the maxima of MI gains of higher orders are shifted to higher wave numbers $k$. The opposite shift direction occurs for the defocusing regime. As compared to the case of Kerr-type media~\cite{Rev22}, this apparently exhibits the stochastic origin of the MI gains. The similar situation takes place for media with the rectangular response function (Fig.~\ref{fig4}(a)). Fig.~\ref{fig4}(b) provides the dependence of the maximum of the main MI gain peak on the nonlocality parameter $\lambda$ for the case when the quintic nonlinearity term dominates ($\chi<1$). Such a dependence ref\/lects the possibility for MI gain maximum to exceed the corresponding value of MI gain for a local random medium in a narrow domain of wave numbers $k$ (see Fig.~\ref{fig2}(a)). At the same time Fig.~\ref{fig4}(b) depicts the closing in of MI gains of dif\/ferent orders as the nonlocality is increased, that is, high nonlocality smooths the f\/luctuations of the modulation amplitude growth. This is valid irrespectively of the response properties of the media regarded and regimes of cubic-quintic nonlinearity. In Fig.~\ref{fig5} we demonstrate MI gains $G_2$, $G_4$ and $G_6$ for the exponential response function for the regimes with diverse signs of nonlinearity coef\/f\/icients than considered above and compare them with the competing cubic-quintic nonlinearity regime. For focusing cubic and quintic nonlinearities ($g_0>0$, $s_0>0$) one observes the same characteristic pattern consisting in the suppression MI gains, at the same time as compared to competing nonlinearity MI gains are less suppressed (Fig.~\ref{fig5}(a)). The opposite situation takes place for the case of defocusing cubic and focusing quintic nonlinearities ($g_0<0$, $s_0>0$). The MI gain peaks are suppressed more than their competing cubic-quintic nonlinearity counterparts (Fig.~\ref{fig5}(b)). In both cases one can see the transition from the initial regime (initially defocusing regime on Fig.~\ref{fig5}(a) and focusing one on Fig.~\ref{fig5}(b)) as one changes the chosen set of signs of nonlinearity coef\/f\/icients.

\section{Conclusion}
In this paper for the phenomenological model supporting two regimes of the wave propagation, applying the linear stability analysis we have investigated the
MI of a homogeneous wave in a~nonlocal non-Kerr medium with cubic-quintic nonlinearity and random parameters.
Adopting a~white-noise model for parameter f\/luctuations, we have obtained the equations
determining the dependence of the MI gain on the
modulation wave number. Nonlocality was proved to
suppress considerably the stochasticity-induced MI growth rate for media with the sign-def\/inite
Fourier images of the response
functions. At the same time, it was shown that for the nonlocal media with the
sign-indef\/inite Fourier images
of the response functions MI gain can exceed that for local media for some wave number intervals. Some attention was paid to the cases of focusing cubic and quintic nonlinearities and defocusing cubic and focusing quintic nonlinearities.

\subsection*{Acknowledgements}

The author is grateful to E.V. Doktorov for global help during research process and preparation of the paper.
I wish to thank the Organizers of the Seventh International Conference ``Symmetry in Nonlinear
Mathematical Physics'' (June 24--30, 2007, Kyiv) and ICTP Of\/f\/ice of External Activities for having given me the opportunity
to talk on this subject as well as for local and travel support.

\pdfbookmark[1]{References}{ref}
\LastPageEnding


\begin{thebibliography}{99}

\footnotesize\itemsep=0pt

\bibitem{Rev1} Kivshar~Yu.S., Luther-Davies~B., Dark optical solitons: physics and applications, {\it Phys. Rep.} {\bf298} (1998), 81--197.

\bibitem{Rev2} Mamyshev P.V., Bosshard C., Stegeman G.I., Generation of a periodic array of dark spatial solitons in the regime of ef\/fective amplif\/ication, {\it J. Opt. Soc. Amer. B Opt. Phys.} {\bf 11} (1994), 1254.


\bibitem{Rev3} Hasewaga A., Observation of self-trapping instability of a plasma cyclotron wave
in a computer experiment, {\it Phys. Rev. Lett.} {\bf24} (1970), 1165--1168.

\bibitem{Rev4} Whitham G.B., Nonlinear dispersive waves, {\it Proc. R. Soc. London A} {\bf283} (1965), 238--261.

\bibitem{Rev5} Benjamin T.B., Feir J.E., The disintegration of wave trains on deep water,
{\it J. Fluid Mech.} {\bf27} (1967), 417--430.

\bibitem{Rev6} Wu B., Niu Q., Landau and dynamical instabilities of the superf\/low
of Bose--Einstein condensates in optical lattices, {\it Phys.
Rev. A} {\bf64} (2001), 061603, 4 pages, \href{http://arxiv.org/abs/cond-mat/0009455}{cond-mat/0009455}.

\bibitem{Rev7} Kevrekidis P.G., Frantzeskakis D.J., Pattern forming
dynamical instabilities of Bose--Einstein condensates,
{\it Modern Phys. Lett. B} {\bf18} (2004), 173--202, \href{http://arxiv.org/abs/cond-mat/0406657}{cond-mat/0406657}.


\bibitem{Rev8} Stegeman G.I., Segev M., Optical spatial solitons and
their interactions: universality and diversity, {\it Science}
{\bf286} (1999), 1518--1522.

\bibitem{Rev9} Tai K., Hasegawa A., Tomita A., Observation of modulational instability
in optical f\/ibers, {\it Phys. Rev. Lett.} {\bf56} (1986), 135--138.

\bibitem{Rev10} Kivshar Yu.S., Agraval G.P., Optical solitons:
from f\/ibers to photonic crystals, Academic Press, San Diego, 2003.

\bibitem{Rev11}
Abdullaev F.Kh., Darmanyan S.A., Garnier J., Modulational
instability of electromagnetic waves in inhomogeneous and in
discrete media, {\it Progr. Opt.} {\bf44} (2002), 303--365.

\bibitem{Rev12} Peccianti M., Conti C., Alberici E., Assanto G., Spatially incoherent modulational instability in a nonlocal medium, {\it Laser Phys. Lett.} {\bf2} (2005), 25--29, \href{http://arxiv.org/abs/physics/0409139}{physics/0409139}.

\bibitem{Rev13} Dreischuh A., Paulus G.G., Zacher F., Grasbon F.,
Walther H., Generation of multiple-charged optical vortex
solitons in a saturable nonlinear medium, {\it Phys. Rev. E}
{\bf60} (1999), 6111--6117.

\bibitem{Rev14} Suter D., Blasberg T., Stabilization of transverse
solitary waves by a nonlocal response of the nonlinear medium,
{\it Phys. Rev. A} {\bf48} (1993), 4583--4587.

\bibitem{Rev15} Perez-Garcia V.M., Konotop V.V., Garcia-Ripoll J.J.,
Dynamics of quasicollapse in nonlinear Schr\"odinger systems with
nonlocal interactions, {\it Phys. Rev. E} {\bf62} (2000), 4300--4308.

\bibitem{Rev16} Turitsyn S.K., Spatial dispersion of nonlinearity and stability of
multidimensional solitons, {\it Teoret. Mat. Fiz.} {\bf64} (1985), 226--232
(English transl.: {\it Theoret. and Math. Phys.} {\bf64} (1985), 797--801).

\bibitem{Rev16a} Bang O., Krolikowski W., Wyller J., Rasmussen J.J., Collapse arrest and soliton stabilization in nonlocal nonlinear media, {\it Phys. Rev. E} {\bf66} (2002), 046619, 5 pages, \href{http://arxiv.org/abs/nlin.PS/0201036}{nlin.PS/0201036}.

\bibitem{Rev17} Dreischuh A., Neshev D., Peterson D.E., Bang O., Krolikowski W.,
Observation of attraction between dark solitons, {\it Phys.
Rev. Lett.} {\bf96} (2006), 043901, 4 pages, \href{http://arxiv.org/abs/physics/0504003}{physics/0504003}.

\bibitem{Rev18} Krolikowski W., Bang O., Rasmussen J.J., Wyller J., Modulational instability in nonlocal nonlinear
Kerr media, {\it Phys. Rev. E} {\bf64} (2001), 016612, 8 pages, \href{http://arxiv.org/abs/nlin.PS/0105049}{nlin.PS/0105049}.

\bibitem{Rev18a} Krolikowski W., Bang O., Nikolov N.I., Neshev D., Wyller J., Rasmussen J.J., Edmundson~D., Modulational instability, solitons and beam propagation in spatially nonlocal nonlinear media, {\it J. Opt. B Quantum Semiclass. Opt.} {\bf6} (2004), S288--S294,
    \href{http://arxiv.org/abs/nlin.PS/0402040}{nlin.PS/0402040}.

\bibitem{Rev19} Roussignol P., Ricard D., Lukasik J., Flytzanis C., New results on optical phase conjugation in semiconductor-doped glasses, {\it J. Opt. Soc. Amer. B Opt. Phys.} {\bf4} (1987), 5.

\bibitem{Rev20} Lederer F., Biehlig W., Bright solitons and light bullets in semiconductor waveguides, {\it Electron. Lett.} {\bf30} (1994), 1871--1872.

\bibitem{Rev21} Lawrence B., Cha M., Torruellas W.E., Stegeman G.I., Etemad S., Baker G., Kajzar F., Measurement of the complex nonlinear refractive index of single crystal p-toluene sulfonate at 1064 nm, {\it Appl. Phys. Lett.} {\bf64} (1994), 2773--2775.


\bibitem{Rev22} Doktorov E.V., Molchan M.A., Modulational instability in nonlocal Kerr-type
media with random parameters, {\it Phys. Rev. A} {\bf75} (2007), 053819, 6 pages, \href{http://arxiv.org/abs/0704.1093}{arXiv:0704.1093}.

\bibitem{Rev22a}  Wyller J., Krolikowski W., Bang O., Rasmussen J.J., Generic features of modulational instability in nonlocal Kerr media, {\it Phys. Rev. E} {\bf66} (2002), 066615, 13 pages, \href{http://arxiv.org/abs/nlin.PS/0105049}{nlin.PS/0105049}.

\bibitem{Rev23} Novikov E.A., Functionals and the random force
method in turbulence theory, {\it Zh. Exp. Teor. Fiz.}
{\bf47} (1964), 1919--1926 (English transl.: {\it Sov. Phys. JETP} {\bf20} (1964), 1290--1295).
\end{thebibliography}
\end{document}